\documentclass[twocolumn,aps,pra,superscriptaddress,showpacs,longbibliography]{revtex4-1} 
\pdfoutput=1
\usepackage{graphicx} 
\usepackage[version=3]{mhchem} 
\usepackage[a4paper,plainpages=false,pdfstartview={XYZ null null 0.75}, colorlinks=true, citecolor=blue,linkcolor=blue]{hyperref}   
\usepackage[usenames,dvipsnames]{color}
\usepackage{amsfonts}
\usepackage{wrapfig}
\usepackage{subfigure}
\usepackage{amsmath}
\usepackage{gensymb}
\usepackage{url}

\hypersetup{
    pdftitle = {Orbital angular momentum in electron diffraction and its use to determine chiral crystal symmetries},
    pdfkeywords = {Vortex beam, electron vortex,electron diffraction, orbital angular momentum, chiral, crystal}
}

\newcommand{\diff}{\ensuremath{\operatorname{d}}\!}

\newcommand{\partieel}[2]{\ensuremath{\frac{\partial {#1}}{\partial {#2}}}}

\newcommand{\braket}[3]{\ensuremath{\left<#1\left|#2\right|#3\right>}}
\newcommand{\nextline}[1][]{\nonumber\\#1&}
\newcommand{\Int}[1]{\ensuremath{\displaystyle{\int}\!\!\diff{#1}\,}}
\newcommand{\Intbep}[3]{\ensuremath{\displaystyle{\int}_{#1}^{#2}\!\!\diff{#3}}\,}
\newcommand{\vt}[1]{\ensuremath{\boldsymbol{#1}}}

\newcommand{\figref}[2][]{Fig. (\ref{#2}#1)}

\begin{document}

\title{Orbital angular momentum in electron diffraction and its use to determine chiral crystal symmetries}

\author{Roeland Juchtmans}
\affiliation{EMAT, University of Antwerp, Groenenborgerlaan 171, 2020 Antwerp, Belgium}

\author{Jo Verbeeck}
\affiliation{EMAT, University of Antwerp, Groenenborgerlaan 171, 2020 Antwerp, Belgium}
\begin{abstract}
In this work we present an alternative way to look at electron diffraction in a transmission electron microscope. 
Instead of writing the scattering amplitude in Fourier space as a set of plane waves, we use the cylindrical Fourier transform to describe the scattering amplitude in a basis of orbital angular momentum (OAM) eigenstates.
We show how working in this framework can be very convenient when investigating, e.g., rotation and screw-axis symmetries.
For the latter we find selection rules on the OAM coefficients that unambiguously reveal the handedness of the screw-axis.
Detecting the OAM coefficients of the scattering amplitude thus offers the possibility to detect the handedness of crystals without the need for dynamical simulations, the thickness of the sample nor the exact crystal structure.
We propose an experimental setup to measure the OAM components where an image of the crystal is taken after inserting a spiral phase plate in the diffraction plane and perform multislice simulations on $\alpha$-quartz to demonstrate how the method indeed reveals the chirality.
The experimental feasibility of the technique is discussed together with its main advantages with respect to chirality determination of screw axes.
The method shows how the use of a spiral phase plate can be extended from a simple phase imaging technique to a tool to measure the local OAM decomposition of an electron wave, widening the field of interest well beyond chiral space group determination.
\end{abstract}
\maketitle

%
%
%
\section{Introduction}
An object is chiral when it is not superposable onto its mirror image by rotating and/or translating it. Two chiral objects that differ solely by mirror operations are called enantiomorphs and are labeled as the right- and the left-handed variant of the enantiomorphic pair. Chirality occurs from the macroscopic scale, e.g., a hand, to the nano-meter scale, e.g., molecules and crystals. In the latter it manifest itself not in the physical properties of the material (e.g., density, melting point) but in the interaction with other chiral objects, such as circularly polarized light or other chiral molecules. This is of great importance in, e.g., the pharmaceutical industry since biological processes mostly are chiral. Enantiomorphic compounds in drugs can have distinct effects on the human body, and it is essential for the drug to be enantiomorphically pure in order to have the desired effect. In crystals the chirality can be seen directly from the crystal's space group lacking an improper symmetry element (mirror, mirror glide plane, inversion center, or roto-inversion center). When on top of that the space group possesses a screw-axis apart from the $2_1$ screw-axis, the space group itself is chiral as well. Namely, mirroring a crystal with a chiral space group, (see Table \ref{TabEnantiomorphicSpaceGroups}), will change the space group of the crystal to the other one in the enantiomorphic pair. \\
\begin{table}[htb]
\centering
  \begin{tabular}[b]{r c r c r c}
  \hline
  \hline
  1.&($\mathrm{P}4_1, \mathrm{P}4_3$) & 5. & $(\mathrm{P}6_1, \mathrm{P}6_5)$ & 9. & $(\mathrm{P}3_1, \mathrm{P}3_2)$ \\
  2.& ($\mathrm{P}4_132, \mathrm{P}4_332$) & 6. & $(\mathrm{P}6_2, \mathrm{P}6_4)$ & 10. &$(\mathrm{P}3_112, \mathrm{P}3_212)$\\
  3.&  ($\mathrm{P}4_122, \mathrm{P}4_322$)& 7. & $(\mathrm{P}6_122, \mathrm{P}6_522)$ & 11. & $(\mathrm{P}3_121, \mathrm{P}3_221)$\\
  4.& ($\mathrm{P}4_12_12, \mathrm{P}4_32_12$) & 8. & $(\mathrm{P}6_222, \mathrm{P}6_422)$ &  &\\
  \hline
  \hline
  \end{tabular}
\caption{The 22 chiral space groups in class II, divided in 11 enantiomorphic space group pairs. Mirroring a crystal belonging to one of these space groups will change the crystal's space group to the other one in the enantiomorphic pair.}
\label{TabEnantiomorphicSpaceGroups}
\end{table}
Determining the handedness of a crystal in a transmission electron microscope is a difficult task.
In conventional transmission electron microscopy (TEM) the crystal is seen as a two-dimensional projection on the viewing screen and one can not distinguish between enantiomorphs from one single image, simply because mirroring the crystal along the projection plane changes the chirality of the material, but not its projection on this plane. It is clear that in determining the chirality, one has to find a way to include the third dimension. The most intuitive way of doing this is by rotating the crystal and taking several zeroth order Laue zone (ZOLZ) diffraction patterns, as proposed by Goodman and Secomb \cite{Goodman1977a}. However, in order to determine the absolute rotation direction with respect to the crystal from the diffraction pattern, one has to perform dynamical simulations, which require knowledge about the thickness of the sample and the structure of the crystal. Taking several diffraction patterns under different angles can be avoided when looking at the so-called higher order Laue zones (HOLZs) in the diffraction pattern \cite{Goodman1977}. These spots contain information about the crystal in the direction perpendicular to the projection plane and are suited to determine the chirality. However, in the kinematical approximation Friedel's law imposes extra symmetries on the Fourier transform of the crystal potential that prevent the chirality from being determined from only one diffraction pattern \cite{Friedel}. Therefore, dynamical scattering is needed where, again, simulations have to be compared with the experiment. An additional difficulty is that not all HOLZ diffraction spots are sensitive to the chirality of the crystal and one has to identify these in order use them to determine the chirality\cite{Johnson1994, Inui2003}.

In previous work \cite{Juchtmans2015} we proposed the use of electron vortex beams to determine the chirality of crystals with a threefold screw-axis. As described theoretically by Nye and Berry \cite{Nye}, these electrons' wavefunction has the form
\begin{align}
\Psi(\vt{r})_m=\psi(r,z)e^{im\phi},\label{Vortices}
\end{align}
with $r$, $\phi$, and $z$ the cylindrical coordinates where the $z$ axis is the propagation axis. A vortex electron is an eigenstate of the angular momentum operator $L_z=-i\hbar\partieel{}{\phi}$ and possesses a well-defined amount of angular momentum of $m\hbar$ per electron, where the number $m$ is called the topological charge of the vortex \cite{Allen}. The first experimental realization of optical vortex beams \cite{Bazhenov1990} inspired active research which led to numerous applications ranging from  nano-manipulation \cite{Luo,He,Friese} and astrophysics \cite{Foo2005,Swartzlander2007,Serabyn2010,Berkhout2008} to telecommunication \cite{andrewsbook,wangterabit,vortexnot}.  Also, electron vortex beams can be created in a transmission electron microscope using a phase plate of stacked graphite layers \cite{Uchida2010}, holographic reconstruction \cite{Verbeeck2010}, lens aberrations \cite{Clark2013}, or the Aharonov-Bohm effect at the tip of a magnetized needle \cite{Beche2013}, and recent research suggests possible applications in electron energy-loss magnetic chiral dichroism experiments \cite{Verbeeck2010}, spin-polarization devices \cite{Karimi2012}, magnetic plasmons \cite{Mohammadi2012} and nano-manipulation \cite{Verbeeck2013}.\\
 Since vortex beams are chiral, it may be expected that they interact differently with enantiomorphic crystals. Indeed, we showed that when focusing a vortex over an odd screw-axis with only one heavy atom in its vicinity, the symmetry of the HOLZs will depend on the chirality of the screw-axis \cite{Juchtmans2015}. This gives a quantitative method to determine the chirality of crystals in transmission electron microscopy, where multiple scattering is not required and dynamical simulations can be avoided. It also demonstrates the use of electron vortex beams and their chiral character in crystallography and shows that modifying the electron probe can be of use to investigate certain properties. The technique, however, is impeded by the extreme sensitivity of the probe position which has to be maintained exactly on top of the screw-axis during the time of acquisition. Additionally, the symmetry argument can only be used to determine the chirality of odd screw axes. For even screw axes, the diffraction pattern still has to be compared with numerical simulations.\\
In this work we propose an alternative way of looking at electron diffraction. Instead of decomposing the scattering amplitude in Fourier space as a set of plane waves, we write the scattering amplitude as a sum of electron vortices described Eq. \eqref{Vortices}. Using this framework, we study electron scattering on chiral crystals with screw-axis symmetry and find that by looking at the OAM coefficients of the scattered wave the chirality of such crystals can be determined without the need for dynamical simulations, knowledge about the thickness of the sample, or the crystal structure. We also propose a way to measure these components in a transmission electron microscope (TEM) and discuss the main advantages and obstacles of the technique. We start with analytical calculations in the first order Born approximation, extend this to include multiple scattering, and verify our findings with multislice simulations. 

\section{Theoretical formulation\label{Theory}}
The scattering amplitude for an incident plane wave electron with wave vector $\vt{k}$ to be scattered by a potential $V(\vt{r})$ to a plane wave with wave vector $\vt{k}'$ is called the scattering amplitude $A(\vt{k}')=A(k'_\perp,\phi_{k'},k'_z)$ with $k'_\perp,\phi_{k'}$, and $k'_z$ the radial, angular and $z$ component in cylindrical coordinates.
Since this function is periodic in the polar coordinate $\phi_{k'}$, it can be expanded into a set of discrete Fourier coefficients \cite{Wang}:
\begin{align}
A(k'_\perp,\phi_{k'},k'_z)=\frac{1}{\sqrt{2\pi}}\sum_m a_m(k'_\perp,k'_z)e^{im\phi}.
\end{align}
In doing so, we write the scattering amplitude as a sum of vortices with a well-defined OAM of $m\hbar$ [see Eq. \eqref{Vortices}]. The OAM coefficients $a_m(k_\perp,k_z)$ are given by
\begin{align}
a_m(k_\perp,k_z)=\frac{1}{\sqrt{2\pi}}\Intbep{0}{2\pi}{\phi_k}A(k_\perp,\phi_k,k_z)e^{-im\phi_k},\label{Coefficientskz2}
\end{align}
and play an important role in the description of the angular momentum of the total scattering amplitude \cite{Cohen1977,Molina2001}.\\
In Appendix \ref{AppA} we show how in the first-order Born approximation the scattering amplitude for an incoming plane wave electron with wave function $\Psi(\vt{r})=e^{ik_0z}$ to scatter to a plane wave $\Psi'(\vt{r})=e^{i\vt{k}'\cdot\vt{r}}$ on a potential $V(\vt{r})$ is given by

\begin{align}
A(\vt{k}')=\sum_{m}\frac{i^{-m}}{\sqrt{2\pi}}V_m(k'_\perp,k'_z-k_0)e^{im\phi_{k'}},\label{CylScatAmpl}
\end{align}
with 
\begin{align}
V_m(k_\perp,k_z)=&\Intbep{-\infty}{+\infty}{z}\Intbep{0}{\infty}{r}\Intbep{0}{2\pi}{\phi}rV(\vt{r})\nextline
\times \frac{1}{\sqrt{2\pi}}J_m(k_{\perp}r)e^{-im\phi}e^{-ik_zz},
\end{align}
the cylindrical Fourier coefficients of the potential, analogous to the polar Fourier coefficients in two dimensions \cite{Wang}. Note that these coefficients depend on the origin of $r$, which defines the point around which the expansion is taken. Shifting the potential with respect to the origin, will result in different coefficients.\\
For a crystal with periodicity in the $z$ direction, they can be rewritten as
\begin{align}
V_{m,n}(k_\perp)&=\frac{1}{P}\Intbep{0}{P}{z}\Intbep{0}{\infty}{r}\Intbep{0}{2\pi}{\phi}rV(\vt{r})\nextline
\times \frac{1}{\sqrt{2\pi}}J_m(k_{\perp}r)e^{-im\phi}e^{-in\frac{2\pi}{P}z}\label{Vmn},
\end{align}
where $P$ is the periodicity in the $z$ direction and $k_z=n\frac{2\pi}{P}$, $n\in\mathbb{N}$. Because of the periodicity, the latter is quantized, which results in a quantized transferred forward momentum $k_0-k'_z$ in Eq. \eqref{CylScatAmpl}.\\
In what follows we consider consider only elastically scattered electrons for which the energy and the magnitude of the wavenumber are conserved during the scattering process. We thus have $k^2_0=k'^2_\perp+k'^2_z$ and can write the scattering amplitude as
\begin{align}
A_n(\phi_{k'})&=\sum_{m}\frac{i^{-m}}{\sqrt{2\pi}}V_{m,n}e^{im\phi_{k'}}\label{Anphi}
\end{align}
with $n=\frac{P}{2\pi}(k'_z-k_0)$ and $k'_\perp=\sqrt{k^2_0-k'^2_z}$. Because of the discretized transverse momentum, the diffraction pattern will consist of discrete rings that can be labeled with the number $n$ and are known in conventional electron diffraction theory as the zeroth order Laue zone $(n=0)$, the first order Laue zone $(n=1)$, and the higher-order Laue zone  $(n>0)$, hereafter referred  to as ZOLZ, FOLZ, and HOLZ, respectively. Note that in this assumption, the ZOLZ would only consist of one single point, while in reality it consists of a region of Bragg spots. The reason for this is that the crystal has a finite size in the $z$ direction, blurring out the reciprocal lattice points of the potential in Fourier space. This effect is mainly visible in the ZOLZs where the Ewald sphere lies relatively flat with respect to the zone axis, but it is less important for the HOLZs where the Ewald sphere cuts the harmonics of the $z$ periodicity under an angle \cite{DeGraef2003}.\\
When a crystal possesses screw-axis symmetry, it is invariant under the combined transformation of rotating it around a point on the screw-axis over a certain angle $\Delta \phi$ while translating it over a distance $\Delta z$ parallel to the rotational axis. Because of the periodicity of the crystal, there are only a limited amount of possibilities for $\Delta \phi$ and $\Delta z$,
\begin{align}
\Delta \phi=\frac{2\pi}{M}\hspace{1cm} \text{and} \hspace{1cm} \Delta z=\frac{lP}{M}, \label{eq:dphidz}
\end{align} 
with $M\in \{2,3,4,6\}$ the order of the rotation, $P$ the period of the crystal along the screw-axis direction, and $l\in\{0,\dots,M-1\}$ the translation along the axis in units of $P/M$. In the Hermann-Mauguin notation of space groups used in Table \ref{TabEnantiomorphicSpaceGroups}, a screw-axis is denoted as $M_l$.\\
In Appendix \ref{AppB} we show that the screw-axis symmetry puts restrictions on the polar Fourier coefficients in Eq. \eqref{Vmn} when the crystal is expanded around a point on the screw-axis. For a crystal with $M_l$ screw-axis symmetry, the nonzero coefficients in the expansion have to suffice,
\begin{align}
m=NM-nl\label{mRestriction}
\end{align}
with $N\in\mathbb{N}$. From Eq. \eqref{Anphi}, it is clear that every $n$-th diffraction ring can be seen as a sum of OAM eigenstates $e^{im\phi_k}$, where the components are given by the coefficient $V_{m,n}$. As a result,  the restriction in Eq. \eqref{mRestriction} will also be reflected in the OAM states contributing to the different Laue zones.\\
As an example, consider a crystal that belongs to space group $P6_2$ for which $M=6$ and $l=2$. In this case the first order Laue zone, $n=1$, will only have OAM coefficients for which $m\in\{\dots,-8,-2,4,10,\dots\}$, while its enantiomorph, belonging to space group $P6_4$ will only have components $m\in\{\dots,-10,-4,2,8,\dots\}$. Here the periodicity at which the OAM components occur is given by $M$.\\
It is clear that if we would be able to measure the OAM components of a HOLZ, we could directly identify the crystal's chirality. Moreover, it is sufficient to determine for a limited amount of coefficients whether they are zero or not, which is generally easier to detect than to measure their actual value. Note that this property arises because of the symmetry and is independent of the crystal structure itself.  
\section{Measuring OAM coefficients of HOLZs numerically}
In order to investigate the OAM decomposition of scattered electrons in more detail, we perform multi-slice simulations with the program STEMSIM \cite{Rosenauer2007}. We study $\alpha$-quartz that belongs to space group $P3_121$ or $P3_221$, where we adopt the structure from Baur \cite{Baur2009}. The two enantiomorphs posses a right- or a left-handed screw-axis for which $M=3$, and $l=1$ ans $l=2$, respectively. In \figref[a]{Fig1} the simulated image is shown for a sample of 20\,nm thickness.\\
In \figref[b]{Fig1} the diffraction pattern is calculated by taking the square of the Fourier transform of the simulated exitwave. For the ZOLZ, the disk of spots in the middle, contains information on the projection of the potential on the plane of view. The circle of spots around the ZOLZ is the FOLZ and contains information of the crystal potential in the direction perpendicular to the plane of view and therefore can be used to determine the chirality of the crystal. Since the FOLZ involves high angle scattering, its signal is considerably weaker than the HOLZ due to the atomic scattering factors.
\begin{figure}[t!]
\includegraphics[width=\columnwidth]{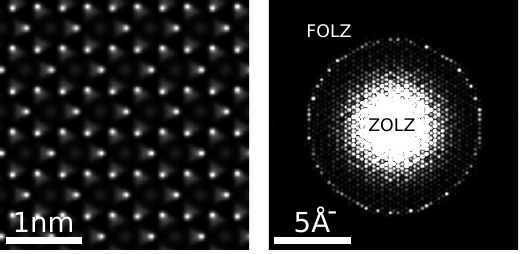}
\caption{Simulated TEM image (Left) and diffraction pattern (Right) of right-handed $\alpha$-quartz with a 300keV electron beam, samplethickness 20\,nm. The circle of Bragg spots in the middle of the diffraction pattern is the zeroth order Laue-zone that contains information about the projection of the potential on the image plane. The ring of spots around the ZOLZ is the first order Laue-zone. These spots contain information of all three dimensions and thus the chirality of the crystal.\label{Fig1}}
\end{figure}

\begin{figure}[t!]
\includegraphics[width=\columnwidth]{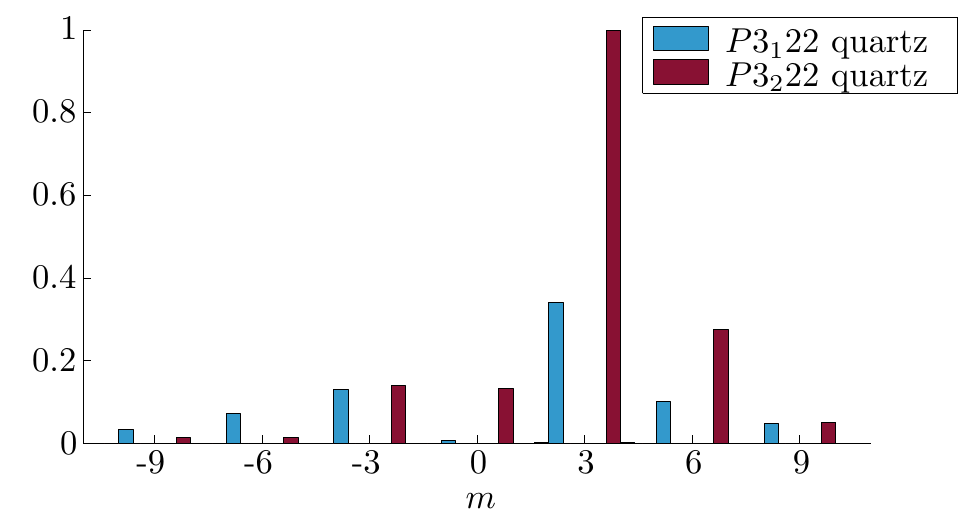}
\caption{Simulated OAM coefficients, $|a_{m,1}(k_\perp)|^2$, $m\in\{-10,10\}$ of the FOLZ in the diffraction pattern of left- (red) and right- (blue) handed $\alpha$-quartz, sample thickness 20\,nm, assuming the optical axis to be centered on top of a screw-axis. Consistent with Eq. \eqref{mRestriction}, for the left-handed enantiomorph only the coefficients with $m\in\{\dots,-2,1,4,\dots\}$ are present, while for the right-handed variant $m\in\{\dots,-4,-1,2,\dots\}$.\label{20nmCoefficients}}
\end{figure}

From the Fourier transform of the simulated exit wave, we can numerically calculate the coefficients $a_{m,n}$ for any $n$-th Laue zone. First all pixels in Fig. \ref{Fig1}(b) are set to zero, except for the ones in a small region around the Laue zone of interest. Then each pixel is multiplied by a factor $\exp(-im\phi)$. The coefficients then are obtained by adding up all pixel values which is the numerical approximation of the integral in \eqref{Coefficientskz2}.\\
The magnitude squared of the coefficients that build up the FOLZ, $|a_{m,1}(k_\perp)|^2$, $m\in\{-10,10\}$  for right- and left-handed $\alpha$-quartz, sample thickness 20\,nm, are shown in Fig. (\ref{20nmCoefficients}).
Consistent with Eq. \eqref{mRestriction}, it can be seen that only the coefficients $a_{m,n}$ for which $m=NM-nl$ appear in the OAM decomposition. In our example for the left-handed $\alpha$-quartz ($l=2, M=3$) and selecting the FOLZ ($n=1$) only the coefficients $m=N.3-1.2\in\{\dots,-5,-2,1,4,\dots\}$ contribute, while for the right-handed enanthiomorph the non-zero coefficients are the ones for which $m\in\{ \dots,-4,-1,2,5\dots\}$.
\section{Measuring OAM coefficients of the HOLZs in a TEM}
\begin{figure}
		\includegraphics[width=1\columnwidth, trim = 0cm 0cm 0cm 1cm, clip]{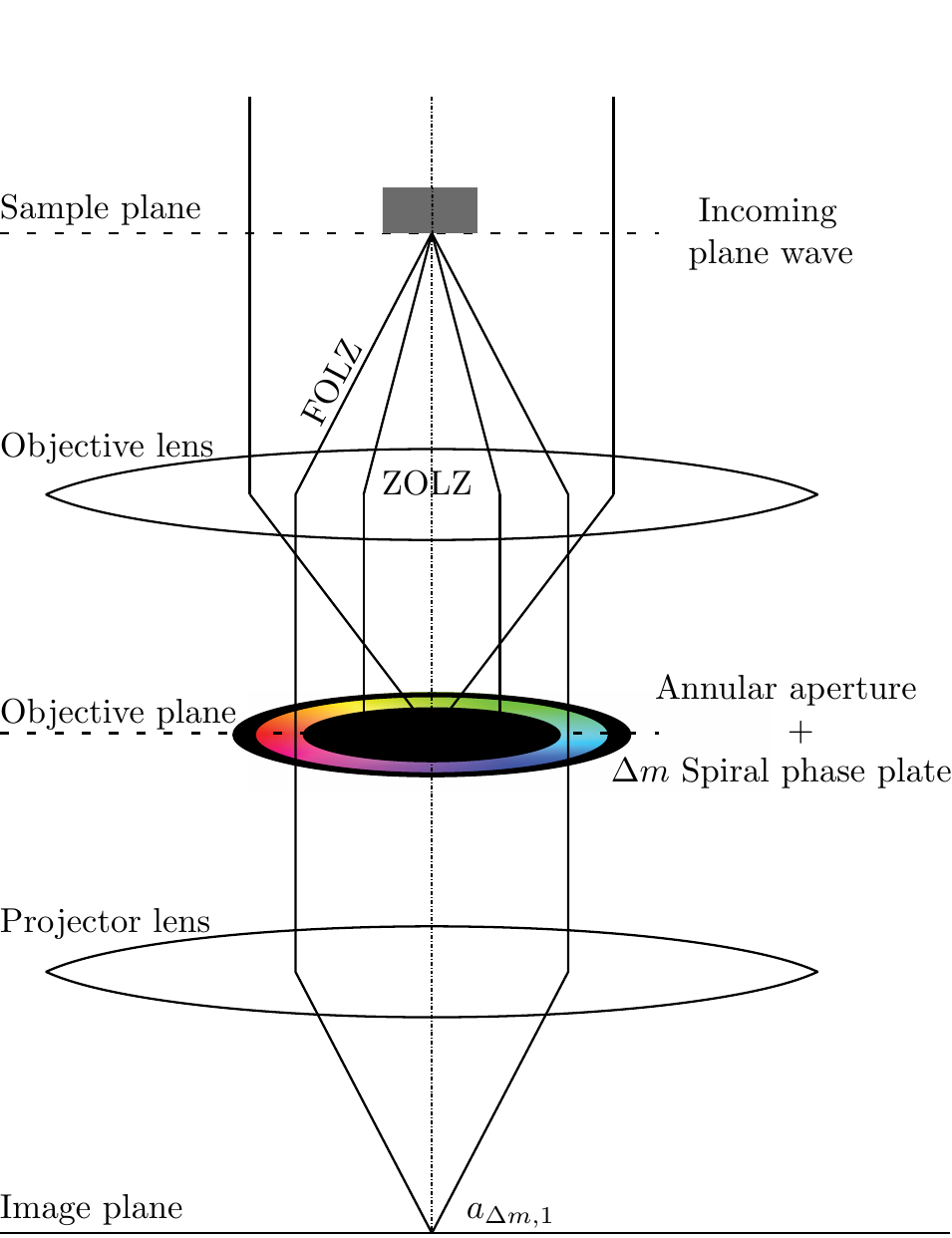}
		\caption{Schematic illustration of experimental setup to measure the OAM coefficients $a_{\Delta m,1}$. An incoming plane wave is scattered on a sample, after which an annular aperture in the objective plane selects only the FOLZ electrons. At the same time a spiral phase plate adds an OAM of $\Delta m$ to the FOLZ. Using a projector lens, we form an image with the resulting electrons where the intensity corresponds to $|a_{\Delta m,1}|^2$ when the decomposition is made around the corresponding point in the sample.\label{figSetup}}
\end{figure}

So far we have looked only at OAM decompositions by means of simulations where we knew both the amplitude and the phase of the scattered wave. In a transmission electron microscope, however, we can not detect the phase of the scattered wave directly, but we can manipulate the scattered wave such that we can detect the OAM coefficients, as we demonstrate below.\\
Let us consider the image created by selecting a HOLZ with an annular aperture in the diffraction plane while inserting a spiral phase plate as proposed in \cite{Blackburn2014}. This experimental setup is schematically shown in Fig. \ref{figSetup}. With a spiral phase plate we mean any phase plate that adds an angular-dependent phase of the form $e^{-i\Delta m\phi_k}$ to the wave in diffraction space. This can be achieved by passing the wave through a material with continuously increasing thickness proportional to the angular coordinate $\phi$ \cite{Uchida2010, Beche2015}, or by exposing the wave to a magnetic monopole-like field induced by the tip of a magnetized needle \cite{Beche2013}. 
These spiral phase plates are not to be confused with the spiral aperture described in \cite{Verbeeck201283}, which only generates a set of vortex beams with different focal points and does not add net OAM to the wave. \\

\begin{figure}[t]
		\includegraphics[width=.64\columnwidth,trim=0cm 0cm 0cm .4cm,clip]{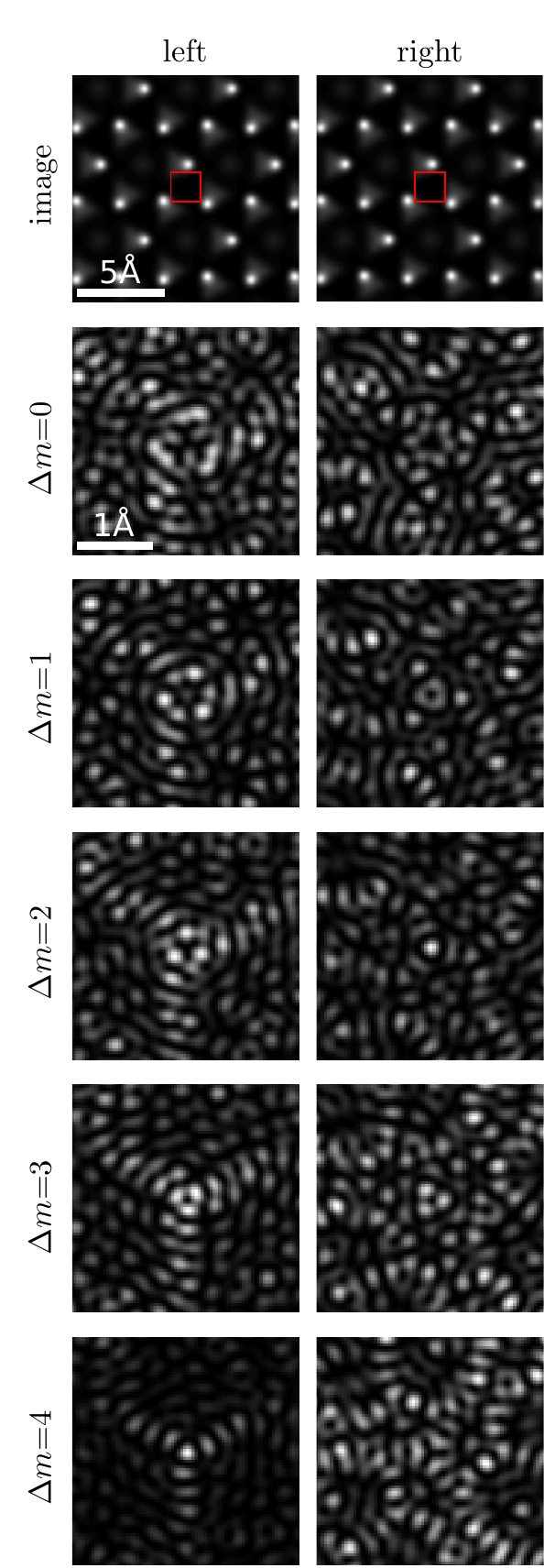}
		\caption{Comparison of images of left- and right-handed 20\,nm thick $\alpha$-quartz. The upper images show the simulated TEM image on which no difference can be seen between the enantiomorphs. The images below are the ones obtained by selecting the FOLZ in Fourier space while changing its OAM by an amount (in units of $\hbar$) $\Delta m \in\{0,1,2,3,4,\}$ of the region indicated by the red square. The upper scale bar applies to the two upper images, while the lower scale bar applies for the OAM -filtered images. Clear differences can be seen between the enantiomorphs of which the appearance of bright spots on the location of the screw-axis can be linked to the handedness of the screw-axis.\label{Zoom}}
		%
	\end{figure}

Mathematically, the resulting wave in the image plane is given by the two dimensional Fourier transform of the scattered wave multiplied by a $\delta$ function selecting the HOLZ and an OAM eigenstate $e^{-i\Delta m\phi_k}$:
\begin{align}
\psi_{\Delta m,n}(r,\phi)=&\Intbep{0}{\infty}{k_\perp}\Intbep{0}{2\pi}{\phi_k}k_\perp A(k_\perp,\phi_k,k_z)e^{-i\Delta m\phi_k}\nextline\times\delta(k_\perp-k^n_\perp)e^{-ik_\perp r \cos(\phi_k-\phi)}.\label{OAMImage}
\end{align}
The point now is that the center of this image ($r=0$) gives us, up to some constants, the expression for the OAM coefficients of the HOLZ [Eq. \eqref{Coefficientskz2}],
\begin{align}
a_{\Delta m,n}\propto\psi_{\Delta m,n}(0),
\end{align}
where again the number $n=\frac{P}{2\pi}(k'_z-k_0)$ and $k'_\perp=\sqrt{k^2_0-k'^2_z}$. This means that the intensity at the center of the image is proportional to $\left|a_{\Delta m,n}\right|^2$. \\
Inserting a phase plate imposing an OAM of $-\Delta m$ to the wave can reveal whether or not the $\Delta m$-th coefficient is present in the decomposition around a certain point in the crystal simply by looking at the intensity of the corresponding point in the image. This is demonstrated in \figref{Zoom}, where a screw-axis lies in the center of the image. For the left-handed $\alpha$-quartz, a bright spot can be seen on the screw-axis for $\Delta m=1$ and $\Delta m=4$, while for right-handed $\alpha$-quartz this is the case for $\Delta m=2$, which is consistent with Eq. \eqref{Coefficientskz2} and with \figref{20nmCoefficients}.\\

\begin{figure}
	\includegraphics[width=\columnwidth]{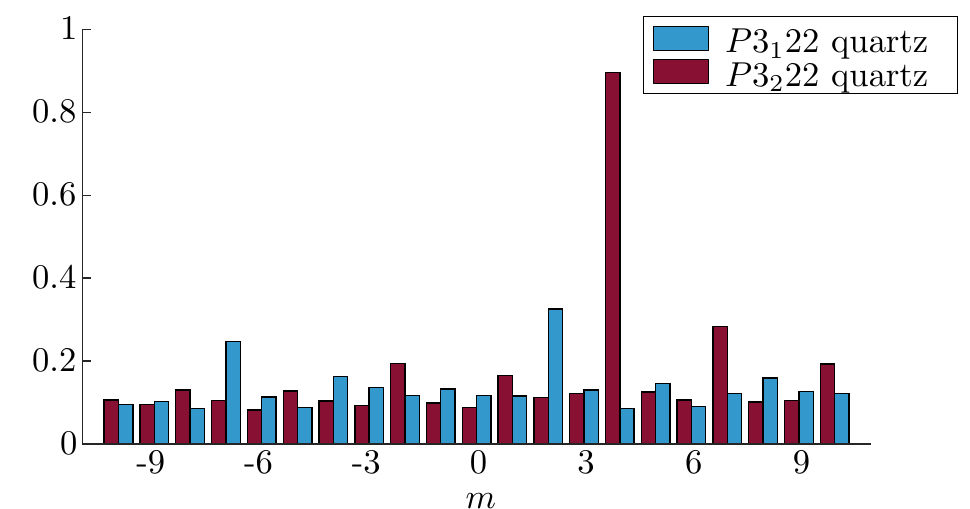}
	\caption{Simulated OAM coefficients, $|a_{m,1}(k_\perp)|^2$, $m\in\{-10,10\}$, of the FOLZ in the diffraction pattern of left- (red) and right- (blue) handed $\alpha$ quartz, sample thickness 20\,nm, including phonon scattering. Previously forbidden components in Fig. 2 now become visible.\label{fig:Phonon}}
\end{figure}

An important note here is that in the theoretical description of the reciprocal space, we have always put the center of our coordinate system on top of a screw-axis. When the crystal is shifted, the screw-axis symmetry with respect to the center of the coordinate system will be broken and the components in the OAM spectrum no longer have to satisfy (\ref{mRestriction}). However, a shift of the sample will only result in a shift of the images. The chirality of the crystal can therefore still be seen from the intensity at the location of the screw axes, which can easily be recognized from the three-fold rotation symmetry that is present at these points. This is further demonstrated in Appendix \ref{AppD} for crystals with space groups $P4_1$ vs. $P4_3$ and $P6_222$ vs. $6_422$. In the $P6_{2(4)}$ case a clear distinction can be made between the sixfold screw-axis on the one hand and the threefold screw axes that are also present, on the other. This shows that, in principle, we can detect all points of screw-axis symmetry in the field of view on the atomic level, making it the ideal method to study screw-axis symmetry behavior on grain boundaries or screw dislocations. \\
Because of dynamical scattering, the diffraction pattern as well as the OAM coefficients depend heavily on the thickness of the sample and the exact structure of the crystal. However, in Appendix \ref{AppC} we show that the restriction on the components only depends on the screw-axis symmetry, independent of the thickness or the structure. This means that, in contrast to previously developed techniques, when determining the chirality from the OAM coefficients of the HOLZs, we no longer require dynamical simulations nor any knowledge about the crystal structure nor the thickness of the sample, as was the case in our previous work using a vortex probe to determine the chirality of a threefold screw-axis \cite{Juchtmans2015}. The big advantage now is that we can determine the chirality of any screw-axis, not only the three-fold, in a qualitative way. Additionally, the major drawback, namely, the exact positioning of the probe on top of a nanometer-sized screw-axis, has been overcome.\\
Because of the symmetry argument and the qualitative nature of the signal, the technique seems very promising. However, there are some challenges for experimental implementation. The first is that it would be preferential to have a method to switch the total amount $\Delta m$ of OAM added in order to be sensitive to multiple OAM coefficients. For now, we only have a magnetic needle that allows us to study two coefficients with opposite OAM. In theory a flipping needle adding OAM $\Delta m=\pm1$ suffices to distinguish between $3_1$ and $3_2$ screw axes, as well as $4_1$ vs. $4_3$ and $6_1$ vs. $6_5$ screw axes, since the difference can be seen by studying the $a_{\pm1,n}$ components. However, this is not the case for the $6_{2(4)}$ screw-axis. Additionally, the experiment would be a lot more conclusive if one would be able to study a series of components giving consistent results. The second and major obstacle is the fact the feature we are looking for in the images, namely, the bright and darker spots, have a size of the order of 0.1\AA\ which requires aberration correction to work up to 86\,mrad or 5$\lambda$. (Currently, 25$\lambda$ is the limit \cite{Kisielowski2008, Erni2009}.) The reason for this is that the images are made out of electrons with high transverse momentum and thus high transverse frequencies.
Besides putting extreme requirements on the resolution, the small size of the features makes the signal very sensitive to phonon scattering, where the atoms deviate from their equilibrium position.
To illustrate this we performed multislice simulations, including phonon scattering, on a right- and a left-handed quartz sample using the experimental Debye-Waller factors from \cite{Tucker2001} obtained at 20 K, the result being shown in Fig. \ref{fig:Phonon}.
We clearly see previously forbidden components appearing that obscure the signal, an effect that increases with growing sample thickness. For thin samples, however, the largest components are still the ones satisfying Eq. \eqref{mRestriction}.”
\\

\section{Conclusion}
By describing electron scattering on chiral crystals with screw-axis symmetry in cylindrical coordinates we were able to write the scattering amplitude as a sum of eigenstates of the orbital angular momentum operator. Using the first-order Born approximation, we found that the screw-axis symmetry puts restrictions on the OAM states building up the HOLZs and that these restrictions are dependent on the order and the handedness of the screw-axis. We verified this with numerical multislice simulations and found this to remain valid even when multiple scattering is dominant. This means that the screw-axis symmetry and the chirality of the crystal can be determined simply by looking at which OAM coefficients are zero and which are not. \\
We propose a method to measure these coefficients by making an image after selecting the FOLZ with an annular aperture in the objective plane, while modifying its OAM by an amount $\Delta m$ using a magnetic needle as described by B\'{e}ch\'{e} \emph{et al}. \cite{Beche2013}. We calculate that the OAM components of the FOLZ then are proportional to the intensity of the rotation symmetric points in the image and show with numerical simulations that the chirality can be seen from bright and dark spots in high-symmetry points in the image.\\
Although the OAM components depend on the thickness of the sample and the exact structure of the crystal, we show theoretically and numerically that the restriction on the components depends only on the screw-axis symmetry, independent of the thickness or the crystal's structure. This means that our proposed method does not require dynamical simulations, nor any knowledge about the crystal structure or the thickness of the sample. In previous work we demonstrated that this was possible for crystals with a threefold screw-axis using electron vortex beams. The new method, however, extends this for all possible screw axes, while the main disadvantages of the vortex technique, namely, the exact positioning of the probe on the screw-axis, is overcome. On top of that, all screw axes in the field of view can be determined at once on the atomic scale, while in the vortex setup we could only determine the chirality of one screw-axis at a time.\\
The main challenge for experimental implementation is the required resolution of the order of 0.1\,\AA\ for a 300-keV probe. Although this lies well within the diffraction wavelength, it lies beyond the resolution of current state-of-the-art microscopes. Reaching these may require technological breakthroughs which pose serious challenges. Also, adding a range of different amounts of OAM to the scattered wave in Fourier space such that we can acquire a series of images with a different OAM filter would be preferential, where currently we can only impose only two values with opposite sign using a single magnetic needle.\\
Finally, this work shows that looking at the OAM decomposition of a scattered wave can reveal extra information about a sample if one understands how the scattering process influences the OAM of the wave. The experimental method to measure the OAM components proposed here would also work without FOLZ filtering, in which case it would be directly applicable in a conventional TEM without the extreme performance which is needed here. Such a method could find use way beyond the field of chiral space group determination.

%
%
%
\section{acknowledgments}
The authors acknowledge support from the FWO (Aspirant Fonds Wetenschappelijk Onderzoek - Vlaanderen), the EU under the Seventh Framework Program (FP7) under a contract for an Integrated Infrastructure Initiative, Reference No. 312483-ESTEEM2 and ERC Starting Grant 278510 VORTEX.
\nocite{DeGraef2003}
\appendix
\section{Diffraction in cylindrical coordinates \label{AppA}}
Consider a plane-wave electron with wave vector $\vt{k}$ scattered on a potential $V(\vt{r})$. In the first-order Born approximation, the scattering amplitude to scatter to a plane wave with wave vector $\vt{k}'$ is given by \cite{DeGraef2003}
\begin{align}
A(\vt{k}')=&\braket{\vt{k}'}{V}{\vt{k}}
\nextline[=] \Int{\vt{r}}e^{-i\vt{k}'\cdot\vt{r}}V(\vt{r})e^{i\vt{k}\cdot\vt{r}}.\label{scatteringamplitude}
\end{align}
In what follows it will be convenient to have an expression for the scattering amplitude in cylindrical coordinates. To calculate this, we first expand the potential in its cylindrical Fourier components \cite{Wang},
\begin{align}
V(\vt{r})=&\Intbep{-\infty}{+\infty}{k_z}\Intbep{0}{\infty}{k_\perp}k_\perp\sum_{m=-\infty}^{\infty} V_m(k_\perp,k_z)
\nextline\times\frac{1}{\sqrt{2\pi}}J_m(k_{\perp}r)e^{im\phi}e^{ik_zz},\label{cylindricalfourier}
\end{align}
with
\begin{align}
V_m(k_\perp,k_z)=&\Intbep{-\infty}{+\infty}{z}\Intbep{0}{\infty}{r}\Intbep{0}{2\pi}{\phi}rV(\vt{r})\nextline
\times \frac{1}{\sqrt{2\pi}}J_m(k_{\perp}r)e^{-im\phi}e^{-ik_zz}.
\end{align}
For the scattering amplitude for an incoming electron along the $z$ axis with wave number $k_0$, substituting Eq. \eqref{cylindricalfourier} into \eqref{scatteringamplitude} gives us 
\begin{align}
A(\vt{k}')=&\Int{\vt{r}}\Intbep{-\infty}{+\infty}{k_z}\Intbep{0}{\infty}{k_\perp}k_\perp\sum_{m=-\infty}^{\infty} V_m(k_\perp,k_z)
\nextline\times\frac{1}{\sqrt{2\pi}}J_m(k_{\perp}r)e^{im\phi}e^{i(k_z+k_0-k'_z) z}e^{-i\vt{k}'_\perp\cdot\vt{r}_\perp}.
\end{align}
Using the Jacobi-Anger identity,
\begin{align}
	e^{i\vt{k}'_\perp\cdot\vt{r}_\perp}&=e^{ik'_\perp r_\perp \cos(\phi-\phi'_{k})}\nonumber\\
	&=\sum_{m=-\infty}^{\infty}i^mJ_m(k'_\perp r)e^{im(\phi-\phi'_{k})},
\end{align}
we get
\begin{align}
A(\vt{k}')=&\frac{1}{\sqrt{2\pi}}\Int{\vt{r}}\Intbep{-\infty}{+\infty}{k_z}\Intbep{0}{\infty}{k_\perp}k_\perp\sum_{m,m'=-\infty}^{\infty} V_m(k_\perp,k_z)
\nextline\times J_m(k_{\perp}r)e^{im\phi}e^{i(k_z+k_0-k'_z) z}i^{-m'}J_{m'}(k_\perp'r)e^{im'(\phi_{k'}-\phi)}\nextline[=]
\frac{1}{\sqrt{2\pi}}\Intbep{-\infty}{+\infty}{k_z}\Intbep{0}{\infty}{k_\perp}k_\perp\sum_{m,m'=-\infty}^{\infty}V_m(k_\perp,k_z)\nextline\times i^{-m'}e^{im'\phi_{k'}}\Intbep{0}{\infty}{r}J_m(k_{\perp}r)J_{m'}(k_\perp'r)\nextline\times\Intbep{0}{2\pi}{\phi}e^{i(m-m')\phi}	\Int{z}e^{i(k_z+k_0-k'_z) z}.
\end{align}
The integrals over $r$, $z$, and $\phi$ give two Dirac $\delta$ functions and a Kronecker $\delta$. The final expression for the scattering amplitude thus becomes
\begin{align}
A(\vt{k}')=&\frac{1}{\sqrt{2\pi}}\Intbep{-\infty}{+\infty}{k_z}\Intbep{0}{\infty}{k_\perp}k_\perp\sum_{m,m'=-\infty}^{\infty} V_m(k_\perp,k_z)
\nextline\times i^{-m'}e^{im'\phi_{k'}}\frac{\delta\left(k_\perp-k'_\perp\right)}{k_\perp}\delta_{m,m'}\delta(k_z+k_0-k'_z)
\nextline[=]\frac{1}{\sqrt{2\pi}}\sum_{m}i^{-m}V_m(k'_\perp,k'_z-k_0)e^{im\phi_{k'}}.\label{CylScatAmp}
\end{align}
Equation \eqref{CylScatAmp} gives the most general form of a scattering amplitude for a plane wave electron to scatter to a plane wave with wave vector $\vt{k}'$ scattered on a potential determined by its cylindrical Fourier coefficients $V_m(k'_\perp,k'_z-k_0)$.\\
\section{Scattering on potential with screw-axis symmetry\label{AppB}}
For a crystal potential with screw-axis symmetry around the origin one has
\begin{align}
V(r,\phi,z)=V(r,\phi+\Delta\phi,z+\Delta z),
\end{align}
where the periodicity of the crystal puts restrictions on $\Delta \phi$ and $\Delta z$,
\begin{align}
\Delta \phi=\frac{2\pi}{M}\hspace{1cm} \text{and} \hspace{1cm} \Delta z=\frac{lP}{M},
\end{align}
with $P$ the period of the crystal along the screw-axis direction, $M\in \{2,3,4,6\}$ the order of the rotation and $l\in\{0,\dots,M-1\}$ the translation along the axis in units of $P/M$.\\A first thing to note is that because of the periodicity in the $z$ direction, the potential can be written as
\begin{align}
V(\vt{r})=&\Intbep{0}{\infty}{k_\perp}k_\perp\sum_{m,n=-\infty}^{\infty} V_{m,n}(k_\perp)
\nextline\times\frac{1}{\sqrt{2\pi}}J_m(k_{\perp}r)e^{im\phi}e^{in\frac{2\pi}{P}z},\label{PolarFourierScrew1}
\end{align}
with
\begin{align}
V_{m,n}(k_\perp)&=\frac{1}{P}\Intbep{0}{P}{z}\Intbep{0}{\infty}{r}\Intbep{0}{2\pi}{\phi}rV(\vt{r})\nextline
\times \frac{1}{\sqrt{2\pi}}J_m(k_{\perp}r)e^{-im\phi}e^{-in\frac{2\pi}{P}z}.
\end{align}
Secondly, the screw-axis symmetry puts restrictions on the contributing components.
\begin{align}
V_{m,n}(k_\perp)&=\frac{1}{P}\Intbep{0}{P}{z}\Intbep{0}{\infty}{r}\Intbep{0}{2\pi}{\phi}rV(r,\phi,z)\nextline
\times \frac{1}{\sqrt{2\pi}}J_m(k_{\perp}r)e^{-im\phi}e^{-in\frac{2\pi}{P}z}
\nextline[=]\frac{1}{P}\Intbep{0}{P}{z}\Intbep{0}{\infty}{r}\Intbep{0}{2\pi}{\phi}rV(r,\phi +\Delta \phi,z+\Delta z)\nextline
\times \frac{1}{\sqrt{2\pi}}J_m(k_{\perp}r)e^{-im\phi}e^{-in\frac{2\pi}{P}z}
\nextline[=]\frac{1}{P}\Intbep{0}{P}{z'}\Intbep{0}{\infty}{r}\Intbep{0}{2\pi}{\phi'}rV(r,\phi',z')\nextline
\times \frac{1}{\sqrt{2\pi}}J_m(k_{\perp}r)e^{-im(\phi'-\Delta \phi)}e^{-in\frac{2\pi}{P}(z'-\Delta z)}
\nextline[=] e^{i(m\Delta \phi+n\frac{2\pi}{P}\Delta z)}V_{m,n}(k_\perp)\label{eq:Vmn}
\end{align}
Therefore the non-zero components have to satisfy the relation
\begin{align}
m\Delta \phi+n\frac{2\pi}{P}\Delta z =&N2\pi
\end{align}
or equivalently
\begin{align}
m=NM-nl\label{Restriction}
\end{align}
with $N\in\mathbb{N}$. Polar Fourier coefficients for which this relations is not fulfilled, will automatically be zero.\\
Knowing this, we can write the scattering amplitude, Eq. \eqref{CylScatAmp}, for a potential with screw-axis symmetry as
\begin{align}
A(\vt{k}')=&\frac{1}{\sqrt{2\pi}}\sum_{m}i^{-m}V_m(k'_\perp,k'_z-k'_0)e^{im\phi_{k'}}
\nextline[=]\frac{i^{nl}}{\sqrt{2\pi}}\sum_{N}i^{-NM}V_{NM-nl,n}(k'_\perp)e^{i(NM-nl)\phi_{k'}}\label{ScatteringAmplitudeScrew}
\end{align}
were the transferred forward momentum, $k'_z-k_0=n\frac{2\pi}{P}$ is discretized because of the periodicity of the potential. For elastically scattered electrons the energy is conserved and the relation $k^2_0=k'^2_\perp+k'^2_z$ has to be satisfied. Since  for a given $k'_z$ or equivalently a number $n$, the variable $k_\perp$ is fixed by the relation $k_\perp=\sqrt{k_z^2-k_z'^2}$ and can be dropped in Eq. \ref{ScatteringAmplitudeScrew}. We then get
\begin{align} 
A_n(\phi_{k'})=\frac{i^{nl}}{\sqrt{2\pi}}\sum_{N}i^{-NM}V_{NM-nl,n}e^{i(NM-nl)\phi_{k'}}\label{ScatteringAmplitudeScrew2}
\end{align}
where
\begin{align}
V_{m,n}=\left.V_{m,n}(k_\perp)\right|_{k_\perp=\sqrt{k_0^2-(n\frac{2\pi}{P}-k_0)^2}}
\end{align}

The discretization of the transverse momentum makes that the diffraction pattern consist out of discrete rings which can be labeled with the number $n$. These rings coincide with the ZOLZ ($n=0$), FOLZ ($n=1$) and higher order Laue zones ($n>0$) known from conventional electron beam diffraction theory. \\
The scattering amplitude for an incident plane wave electron with wave vector $\vt{k}$ to be scattered by a potential $V(\vt{r})$ to a plane wave with wave vector $\vt{k}'$ is called the scattering amplitude $A(\vt{k}')=A(k'_\perp,\phi_{k'},k'_z)$ with $k'_\perp,\phi_{k'}$ and $k'_z$ the radial, angular and $z$ component in cylindrical coordinates.
Since this function is periodic in the polar coordinate $\phi_{k'}$, it can be expended into a set of discrete Fourier coefficients \cite{Wang}.

We can write the scattering amplitude in cylindrical coordinates  $A(\vt{k}')=A(k'_\perp,\phi_{k'},k'_z)$ with $k'_\perp,\phi_{k'}$ and $k'_z$ the radial, angular and $z$ component.
Since this function is periodic in the polar coordinate $\phi_{k'}$, it can be expended into a set of discrete Fourier coefficients \cite{Wang}.
\begin{align}
A(k'_\perp,\phi_{k'},k'_z)=\frac{1}{\sqrt{2\pi}}\sum_m a_m(k'_\perp,k'_z)e^{im\phi}.
\end{align}
In doing so, we write the scattering amplitude as a sum of vortices with OAM $m\hbar$, see Eq. \eqref{Vortices}. The OAM coefficients $a_m(k_\perp,k_z)$ are given by
\begin{align}
a_m(k_\perp,k_z)=\frac{1}{\sqrt{2\pi}}\Intbep{0}{2\pi}{\phi_k}A(k_\perp,\phi_k,k_z)e^{-im\phi_k}.\label{Coefficientskz}
\end{align}
The use of cylindrical coordinates makes it clear how exactly the scattering amplitude is written in terms of OAM eigenstates. From Eq. \eqref{CylScatAmp} it follows that these are proportional to the cylindrical expansion coefficients
\begin{align}
a_m(k'_\perp,k'_z)=i^{-m}V_m(k'_\perp,k'_z-k'_0).
\end{align}
For a crystal with screw-axis symmetry we thus get, Eq. \eqref{Restriction} and \eqref{ScatteringAmplitudeScrew2}
\begin{align}
a_{m,n}=
\begin{cases}i^{-m}V_{m,n}&m=NM-nl\\ \\
0 &\text{otherwise}\end{cases}\label{Coefficients}
\end{align}
As an example consider a crystal that belongs to space group $P6_2$, for which $M=6$ and $l=2$. In this case the OAM decomposition of the first order Laue zone, $n=1$, will only have components for which $m\in\{\dots,-8,-2,4,10,\dots\}$, while its enantiomorph, belonging to space group $P6_4$ will only have components $m\in\{\dots,-10,-4,2,8,\dots\}$. $M$ allways gives the periodicity of the occuring components, while $-l$, in the FOLZ-case, is given by the largest negative $m$-component contributing.\\

\section{Examples of crystals with other screw-axis symmetries \label{AppD}}
\begin{figure}[h!]
\includegraphics[width=\columnwidth]{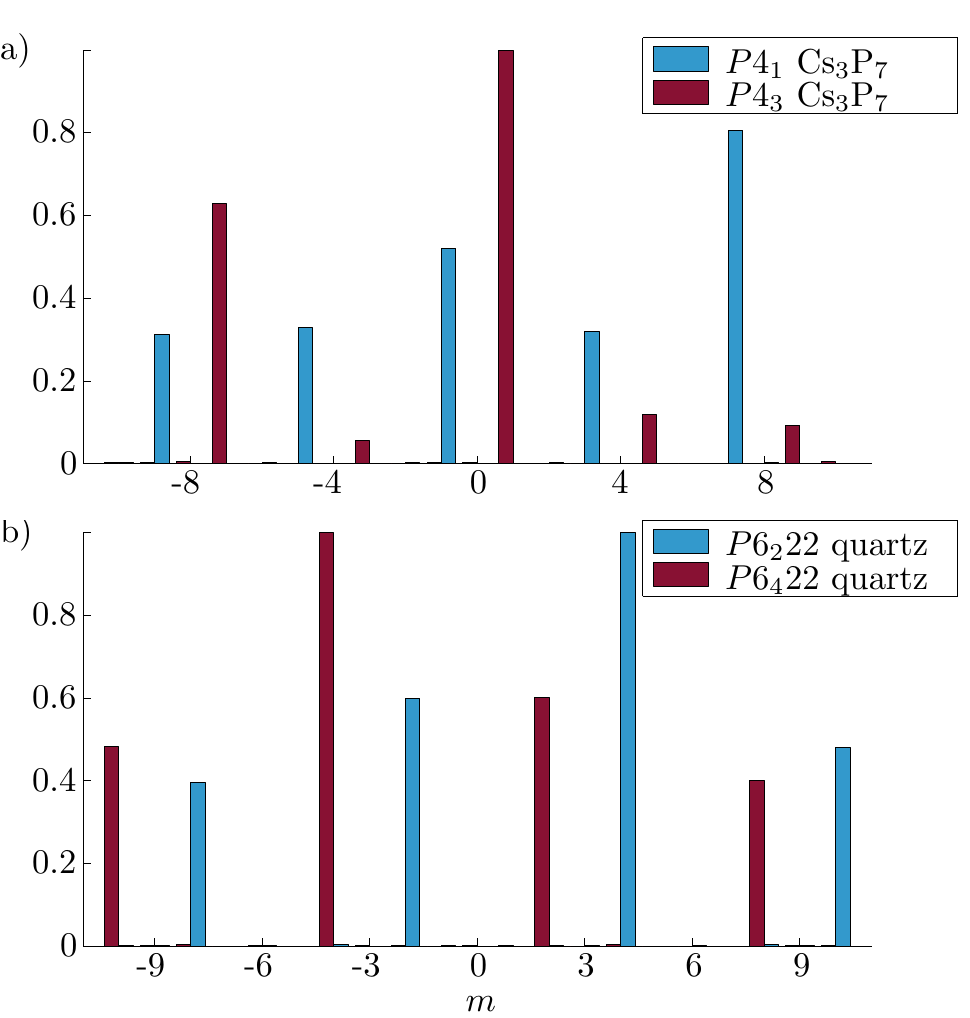}
\caption{Simulated OAM coefficients, $|a_{m,1}(k_\perp)|^2$, $m\in\{-10,10\}$, of the FOLZ in the diffraction pattern of left- (red) and right- (blue) handed (a) Cs$_3$P$_7$ and (b) $\beta$ quartz.Sample thickness is 20\,nm and 200\,nm, respectively. Again, relation \eqref{Coefficients} is fulfilled.\label{CsP}}
\end{figure}

So far we only looked at $\alpha$ quartz with a threefold screw-axis. Of course relation \eqref{Coefficients} puts restrictions on the OAM components of HOLZs of diffraction patterns of crystals belonging to all chiral space groups in Table \ref{TabEnantiomorphicSpaceGroups}. This is demonstrated in \figref{CsP} where we look at the OAM decomposition of $P6_222$ and $P6_422$ $\beta$ quartz \cite{Jacek2006} with respect to the sixfold screw-axis and $P4_1$ and $P4_3$ $\alpha$-tricesium heptaphosphide \cite{Meyer1987}. It is clear that in all cases, indeed, only the OAM components satisfying relation \eqref{Coefficients} are present in the FOLZ. \\

\begin{figure}[h!]
\includegraphics[width=.779\columnwidth]{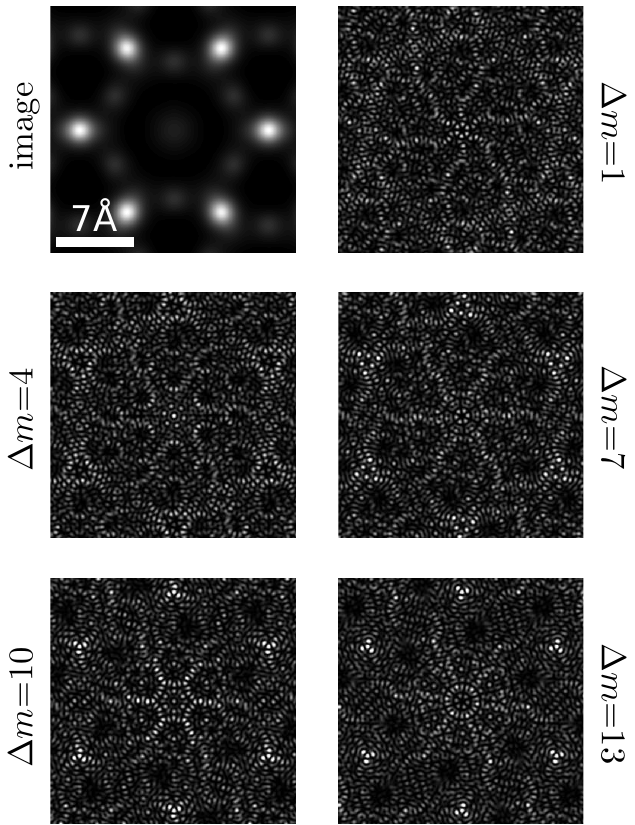}
\caption{TEM image and FOLZ-filtered, OAM modified images of right-handed $\beta$ quartz, $\Delta m\in\{1,4,7,10,13\}$  \label{beta}. From the bright spots a $6_2$ screw-axis in the center and six $3_1$ screw axes at the edges can be identified. Scale bar applies to all subfigures. \label{fig6}}
\end{figure}

\begin{figure}[h!]
\includegraphics[width=.779\columnwidth]{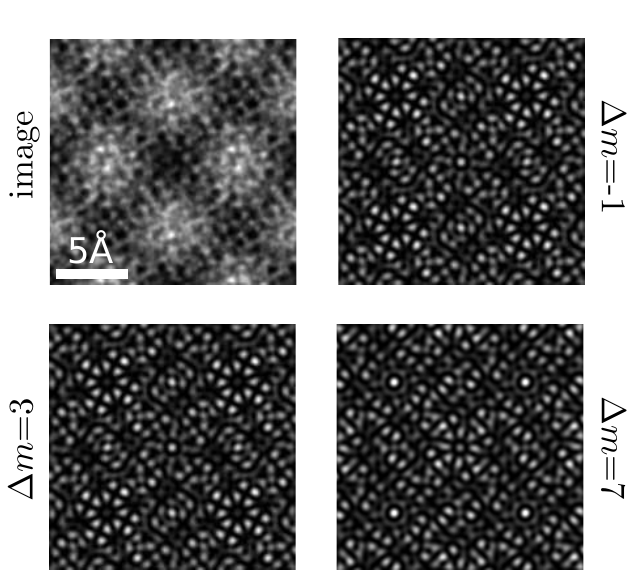}
\caption{TEM image and FOLZ-filtered, OAM modified images of right-handed Cs$_3$P$_7$, $\Delta m\in\{-1,3,7\}$  \label{beta}. From this a $4_1$ screw-axis in the center and in the corners can be identified. As a result of the relatively large periodicity in the direction of the screw-axis, the FOLZ will have a smaller radius, which results in bigger features in the image of the order of $0.5$\,\AA. The scale bar applies to all subfigures. \label{Cs}}
\end{figure}

A big advantage of our technique is that one can identify different screw axes on one series of images, as can be seen in \figref{fig6}. Here we look at the FOLZ-filtered image of $P6_222$ $\beta$-quartz with different $\Delta m$. As expected from \figref[b]{CsP}, we can identify a $6_2$ screw-axis from the bright spots at $\Delta m\in\{4,10\}$. However, at the same time we can see bright spots in the threefold symmetric points at $\Delta m=\{1,4,7,10,12\}$ indicating a $3_1$ screw-axis, which also is present in space group $P6_222$. The technique is thus capable of determining all screw axes and their handedness on the atomic scale in the entire field of view.\\
In Fig. \ref{Cs} we give the FOLZ-filtered images for tricesium heptaphosphide. Because of the relatively large periodicity in the $z$ direction, the FOLZ will have a smaller radius. This means that the electrons that make up the image will have a smaller transverse frequency, which is why the features now are of the order 0.5 \AA\ wide, a factor of 5 bigger than previous examples. Lowering the acceleration voltage will also result in a smaller radius for the FOLZ, having the same effect on the size of the spots we are trying to distinguish. However, doing this generally results in a lower resolution of the microscope. 
\section{Effect of multiple scattering\label{AppC}}
\begin{figure}[h!]
\includegraphics[width=\columnwidth]{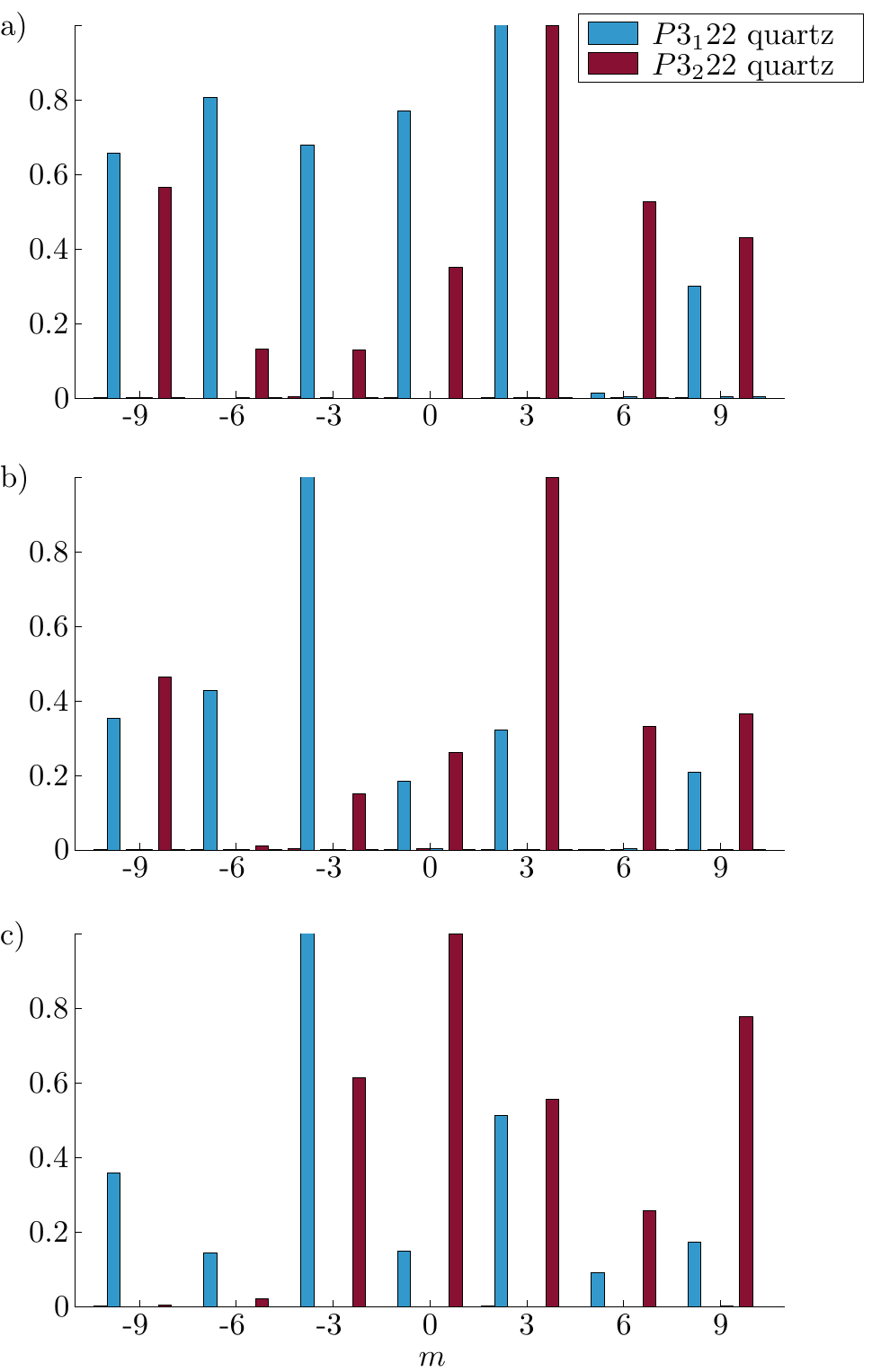}
\caption{Simulated OAM coefficients, $|a_{m,1}(k_\perp)|^2$, $m\in\{-10,10\}$, of the FOLZ in the diffraction pattern of left- (red) and right- (blue) handed $\alpha$ quartz for sample thicknesses (a) 50, (b) 100 and (c) 150\,nm. Although the coefficients depend on the thickness of the sample, they will still obey relation \eqref{Coefficients}, even when multiple scattering is dominant.\label{ThickCoefficients}}
\end{figure}

Of course, electron scattering is mostly dynamical, making the FOLZ and its OAM spectrum dependent on the thickness of the crystal. The latter can be seen in \figref{ThickCoefficients}, where the OAM spectrum is shown for left- and right-handed $\alpha$ quartz for thicknesses ranging from 50\,nm to 150\,nm. However, from this figure it appears that all forbidden coefficients in Eq. \eqref{Coefficients} remain forbidden even when multiple scattering is dominant, independent of the thickness of the sample.\\

\begin{figure}[t]
	\includegraphics[width=\columnwidth]{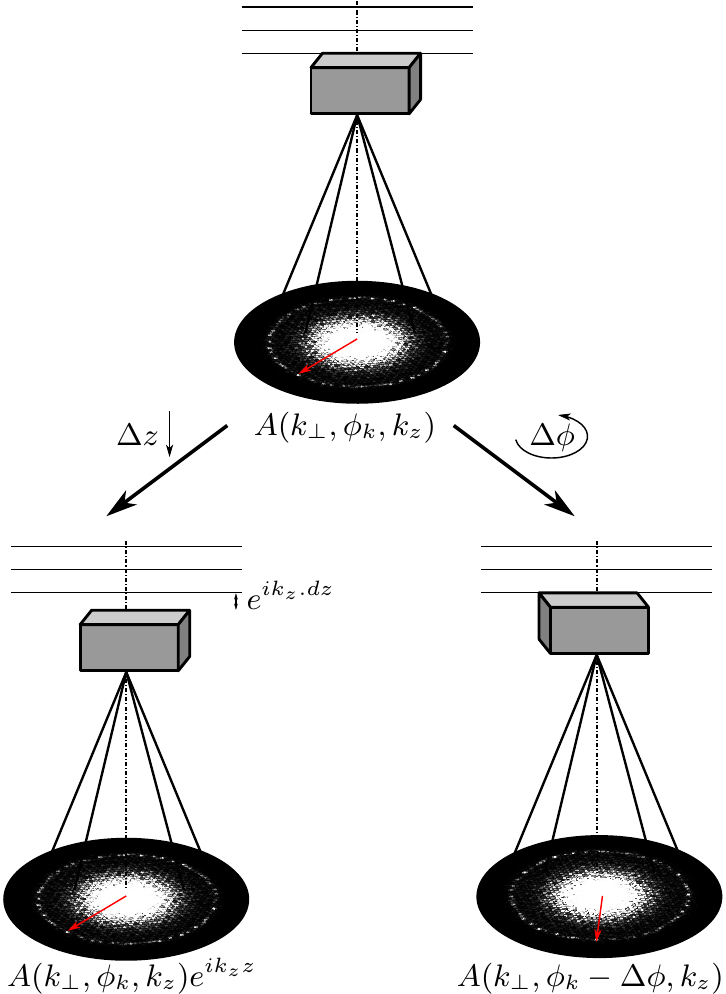}
	\caption{Sketch of the effect of translating a sample along the beam direction and a rotation of the sample around the optical axis. The first adds an extra phase $e^{ik_zz}$ to the scattering, while the second results in a rotation of the scattering amplitude.\label{fig:translationrotation}}
\end{figure}

In order to understand this, we first look at the effect of a translation in the $z$ direction and a rotation around the optical axis of the incoming beam on the scattering amplitude, as illustrated in \figref{fig:translationrotation}. The effect of a translation of the sample along the $z$ direction, the propagation direction of the incoming electron, can be fully incorporated simply by adding a phase $e^{ik_0z}$ to the incoming electron. As an effect of this, the scattering amplitude of a shifted potential $V'(r,\phi,z)=V(r,\phi,z+\Delta z)$, also gets this extra phase such that the scattering amplitude is written as
\begin{align}
A'(k_\perp, \phi_k, k_z)=A(k_\perp,\phi_k,k_z)e^{ik_0.\Delta z}.
\end{align}
Since the incoming electron wave is rotationally invariant, rotating the sample simple results in a rotated scattering amplitude. This means that, when rotating the sample along the optical axis over an angle $\Delta \phi$, the scattering amplitude now becomes
\begin{align}
A'(k_\perp, \phi_k, k_z)=A(k_\perp,\phi_k +\Delta \phi,k_z).
\end{align}

When combining a rotation of the sample with a translation in the $z$ direction, $V'(\vt{r})=V(r,\phi+\Delta\phi,z+\Delta z)$, we thus get for the scattering amplitude

\begin{align}
A'(k_\perp,\phi_k,k_z)&=A(k_\perp,\phi_k+\Delta \phi_k,z)e^{ik_z \Delta z}\label{eq:rotA}.
\end{align}
It is important to note that the treatment includes full dynamical scattering as no kinematical approximation is made.

Let us now consider the special case of a crystal with screw-axis symmetry for which the potential has the property
\begin{align}
V(r,\phi,z)=V(r,\phi+\Delta\phi,z+\Delta z)
\end{align}
for $\Delta \phi$ and $\Delta z$, as before
\begin{align}
\Delta \phi=\frac{2\pi}{M}\hspace{1cm} \text{and} \hspace{1cm} \Delta z=\frac{lP}{M}, \label{eq:dphidz2}
\end{align} 
with $M\in \{2,3,4,6\}$ the order of the rotation, $P$ the period of the crystal along the screw-axis direction, and $l\in\{0,\dots,M-1\}$ the translation along the $z$ axis in units of $P/M$. \\
Since the crystal potential is invariant under this transformation, the scattering amplitude has to be so as well. This means that in Eq. \eqref{eq:rotA} $A'(k_\perp,\phi_k,k_z)=A(k_\perp,\phi_k,k_z)$ and thus
\begin{align}
A(k_\perp,\phi_k,k_z)&=A(k_\perp,\phi_k+\Delta \phi_k,z)e^{ik_z \Delta z}.\label{eq:Arot}
\end{align}
Also, in the dynamical approximation, conservation of energy is assumed, such that we can rewrite the scattering amplitude as $A_n(\phi)$ with $n=\frac{P}{2\pi}(k'_z-k_0)$, $k'_\perp=\sqrt{k^2_0-k'^2_z}$. In this notation the Eq. \eqref{eq:Arot} becomes
\begin{align}
A_n(\phi_k)&= A_n\left( \phi_k+\frac{2\pi}{M}\right) e^{i\frac{n 2\pi}{M}l}.
\end{align}
As before, we can calculate the OAM expansion coefficients for the scattering amplitude using the formula
\begin{align}
a_{m,n}&= \frac{1}{2\pi}\Intbep{0}{2\pi}{\phi_k}A_n(\phi_k)e^{-im\phi_k}.
\end{align}
Filling in relation \eqref{eq:Arot}, eventually gives

\begin{align}
a_{m,n}&= \frac{1}{2\pi}\Intbep{0}{2\pi}{\phi_k}A_n\left( \phi_k+\frac{2\pi}{M}\right) e^{i\frac{n 2\pi}{M}l}e^{-im\phi_k}\nonumber\\
&= \frac{1}{2\pi}\Intbep{0}{2\pi}{\phi_{k'}}A_n\left( \phi_{k'}\right)  e^{i\frac{n 2\pi}{M}l}e^{-im\left( \phi_{k'}-\frac{2\pi}{M}\right) }\\
&=a_{m,n}e^{i(\frac{n 2\pi}{M}l+m\frac{2\pi}{M})}
\end{align}
Therefore the non-zero components have to satisfy the same relation as before
\begin{align}
m\Delta \phi+n\frac{2\pi}{P}\Delta z =&N2\pi,
\end{align}
or equivalently,
\begin{align}
m=NM-nl.
\end{align}
Therefore the nonzero OAM components of the scattering amplitude have to satisfy the relation \eqref{mRestriction}, and in the case of dynamical scattering also.

\end{document}